\documentclass[12pt]{article}
\pdfoutput=1

\usepackage{amsmath,amssymb,amsfonts,graphicx,amsfonts}
\usepackage{textcomp,mathcomp}
\usepackage{color,xcolor}
\usepackage[hidelinks]{hyperref}
\usepackage{ascmac}
\usepackage{physics}

\allowdisplaybreaks[4]

\date{}
\begin{document}
\begin{flushright}
\today\\
\end{flushright}

\vspace{0.1cm}

\begin{center}

 {\Large  Color Confinement and Random Matrices}\\
 \vspace{5mm}
  {\large --- A  random walk down group manifold toward Casimir scaling ---}\\ 

\end{center}
\vspace{0.1cm}
\vspace{0.1cm}
\begin{center}

Georg Bergner$^{a}$, Vaibhav Gautam$^{b,c}$, and Masanori Hanada$^{b}$

\end{center}
\vspace{0.3cm}

\begin{center}

{\small

$^a$ University of Jena, Institute for Theoretical Physics\\
Max-Wien-Platz 1, D-07743 Jena, Germany\\
\vspace{3mm}
$^b$ School of Mathematical Sciences, Queen Mary University of London\\
Mile End Road, London, E1 4NS, United Kingdom\\
\vspace{3mm}
$^c$ 
Department of Mathematics, University of Surrey\\
Guildford, Surrey, GU2 7XH, United Kingdom\\

}
\end{center}

\vspace{1.5cm}

\begin{center}
  {\bf Abstract}
\end{center}

We explain the microscopic origin of linear confinement potential with the Casimir scaling in generic confining gauge theories. 
In the low-temperature regime of confining gauge theories such as QCD, Polyakov lines are slowly varying Haar random modulo exponentially small corrections with respect to the inverse temperature, as shown by one of the authors (M.~H.) and Watanabe. With exact Haar randomness, computation of the two-point correlator of Polyakov loops reduces to the problem of random walk on group manifold. Linear confinement potential with approximate Casimir scaling except at short distances follows naturally from slowly varying Haar randomness. With exponentially small corrections to Haar randomness, string breaking and loss of Casimir scaling at long distance follow. Hence we obtain the Casimir scaling which is only approximate and holds only at intermediate distance, which is precisely needed to explain the results of lattice simulations. For $(1+1)$-dimensional theories, there is a simplification that admits the Casimir scaling at short distances as well.

\newpage
\tableofcontents

\section{Introduction}
\hspace{0.51cm}
In the low-temperature regime of confining gauge theory, the Polyakov line must be slowly varying Haar random up to small corrections and gauge transformations~\cite{Hanada:2023rlk,Hanada:2023krw,Hanada:2020uvt}. We show how linear confinement potential with Casimir scaling~\cite{Ambjorn:1984mb,Ambjorn:1984dp,DelDebbio:1995gc,Bali:2000un,Deldar:1999vi} follows from this fact.\footnote{
In this paper, we consider the Casimir scaling of string tension in the confined phase. The Casimir scaling was also studied for the one-point function of the Polyakov loop in the deconfined phase~\cite{Gupta:2007ax,Gupta:2006qm,Mykkanen:2012ri}. 
} 
The property of a Haar random distribution is explained below in more detail, repeating some of the arguments in Ref.~\cite{Hanada:2023rlk}. It is not based on a strong coupling limit in lattice gauge theory but rather a universal property that emerges for the Polyakov line at a certain range of energy scales. 

\begin{figure}[htbp]
\begin{center}
\scalebox{0.25}{
\includegraphics{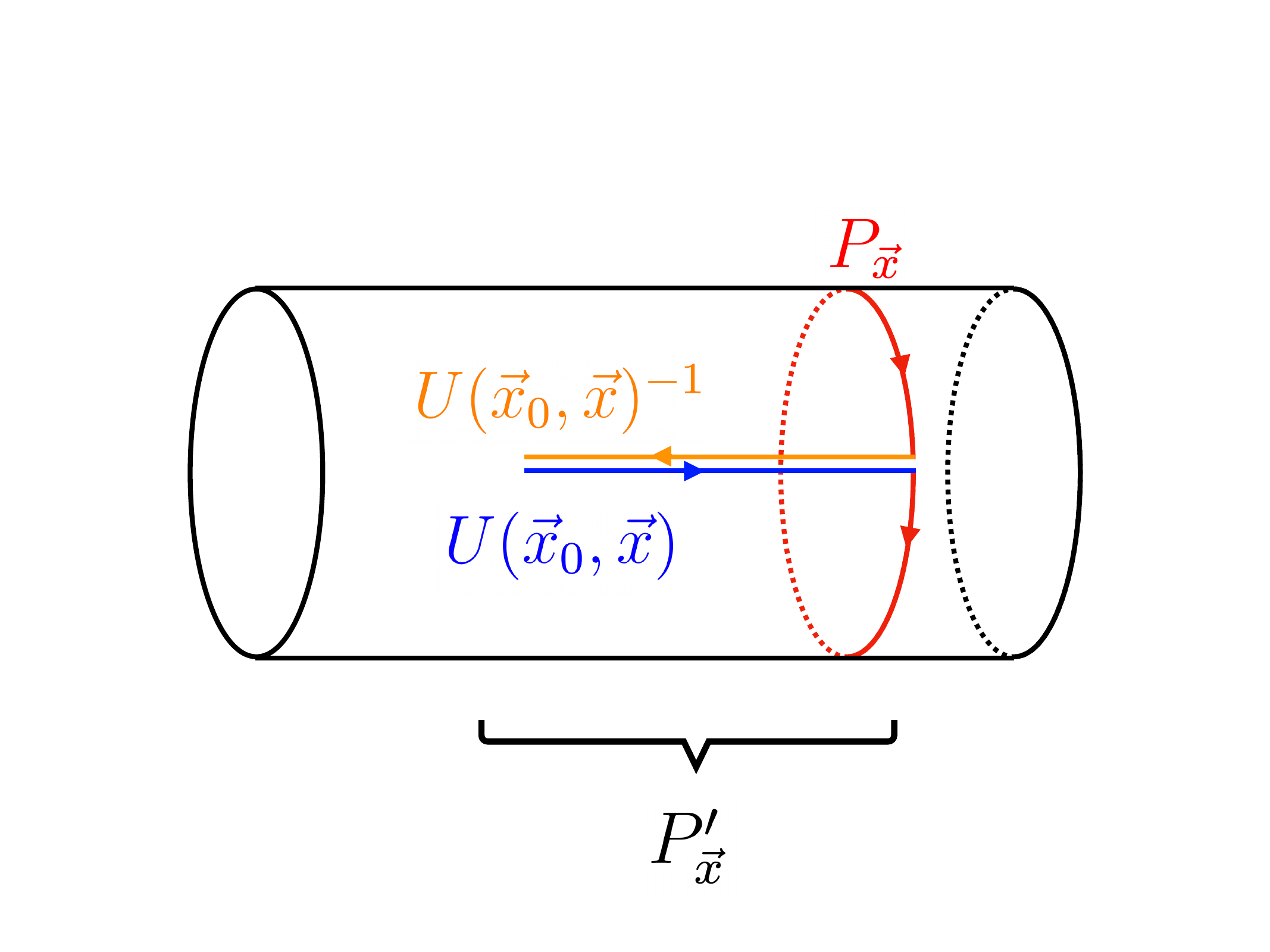}}
\end{center}
\caption{Illustration of
$P'_{\vec{x}}
\equiv
U(\vec{x}_0,\vec{x})
\cdot
P_{\vec{x}}
\cdot
(U(\vec{x}_0,\vec{x}))^{-1}$ defined in \eqref{eq:Pprime} in terms of the path on compact space.
}\label{fig:Figure_P_prime}
\end{figure}

For concreteness, we consider SU($N$) gauge theory, although the same argument applies to any gauge group and any matter content. We quantize this theory via Euclidean path integral at temperature $T=\beta^{-1}$. At each spatial point $\vec{x}$, we can define Polyakov line $P_{\vec{x}}$ as  
\begin{align}
P_{\vec{x}}=\textrm{Path\ ordering}\left[
e^{i\int_0^\beta dt A_t (t,\vec{x})}
\right]\, . 
\end{align}
We choose a reference point $\vec{x}_0$ and define the Wilson line connecting these two points, 
\begin{align}
U(\vec{x}_0,{\vec{x}})=\textrm{Path\ ordering}\left[
e^{i\int_C d\vec{x}\vec{A}}
\right]\, ,  
\end{align}
where $C$ is a contour connecting $\vec{x}_0$ and $\vec{x}$ at fixed Euclidean time $t=0$.
We claim that $P'_{\vec{x}}$ defined by 
\begin{align}
P'_{\vec{x}}
\equiv
U(\vec{x}_0,\vec{x})
\cdot
P_{\vec{x}}
\cdot
U(\vec{x}_0,\vec{x})^{-1}
\label{eq:Pprime}
\end{align}
(see Fig.~\ref{fig:Figure_P_prime}) must be close to slowly varying Haar random, after appropriate smearing of gauge field that will be explained in Sec.~\ref{sec:renormalization}. Namely, statistically, $P'_{\vec{x}}$ is Haar random at each $\vec{x}$, and it is slowly varying as a function of $\vec{x}$ in each typical field configuration that has a large weight in the path integral. 

A different way to express the slowly varying Haar randomness is to consider a change along a straight line $\vec{x}=\vec{x}_0+L\hat{u}$ with some unit vector $\hat{u}$ and length $L>0$. The change of $P_{\vec{x}}$ is a smooth analog of random walk on the group manifold.\footnote{An earlier application of a random walk on group manifold to confinement can be found in Refs.~\cite{Brzoska:2004pi,Arcioni:2005iq,Buividovich:2006yj,Buividovich:2007xh}.
These references introduced a random walk based on phenomenological observation from lattice simulations.
We provide a more complete picture based on the theoretical analysis of gauge theory.
}
For each typical field configuration in path integral, we can write the change of Polyakov line as 
\begin{align}
P_{\vec{x}}^{\prime -1}
P'_{\vec{x}+L\hat{u}}
=
\textrm{Path\ ordering}\left[
e^{i\sum_\alpha \int_0^L dL' v_\alpha(L')T_\alpha}
\right]\, , 
\label{eq:path-ordering}
\end{align}
where $v_\alpha(L)$ ($\alpha=1,2,\cdots,N^2-1$) are smooth functions and $T_\alpha$ are the generators of SU($N$).
This `velocity' $v_\alpha(L)$ gradually and randomly changes. 
We can write a similar expression for any representation r:
\begin{align}
R^{\rm (r)}
\left(
P_{\vec{x}}^{\prime -1}
P'_{\vec{x}+L\hat{u}}
\right)
=
\textrm{Path\ ordering}\left[
e^{i\sum_\alpha \int_0^L dL' v_\alpha(L')T^{\rm (r)}_\alpha}
\right]\, . 
\label{eq:path-ordering-r}
\end{align}

The main part of this paper explains how the Casimir scaling of string tension follows from this random-walk nature. 
In the rest of this introduction section, we will define the Casimir scaling and explain how the computation of string tension is related to the random-walk nature of the Polyakov line. In Sec.~\ref{sec:Hanada-Watanabe-review}, we review the previous work and explain why the Polyakov line has such a property. In Sec.~\ref{sec:epsilon-estimate}, we estimate the rate of change (i.e., the meaning of `slow') based on a lattice regularization. Note that slowly varying Haar randomness does not apply to the abelianized regime where $\mathrm{SU}(N)$ breaks to $\mathrm{U}(1)^{N-1}$ because the correction to slowly varying Haar randomness is too large. Ref.~\cite{Poppitz:2017ivi} confirmed that the Casimir scaling does not hold in such a theory. 

\subsubsection*{Casimir scaling}
We normalize generators of the Lie algebra $T_{\alpha}$ ($\alpha=1,2,\cdots,N^2-1$) in the fundamental representation as 
\begin{align}
\mathrm{Tr}(T_\alpha T_\beta)=2\delta_{\alpha\beta}\, .
\end{align}
For SU(2) and SU(3), we can use Pauli matrices and Gell-Mann matrices, respectively. 
For irreducible representation r, we define the quadratic Casimir invariant $C_{\rm r}$ by 
\begin{align}
\sum_\alpha \left(T^{\rm (r)}_\alpha\right)^2 = 4C_{\rm r}\cdot\textbf{1}\, , 
\end{align}
where $\textbf{1}$ is the identity. 
For spin-$j$ representation of SU(2), $C_{{\rm spin}\mathchar`-j}=j(j+1)$. 
For SU(3), $C_{\rm fund}=\frac{4}{3}$, $C_{\rm adj}=3$, $C_{\rm 2\mathchar`-sym}=\frac{10}{3}$, $C_{\rm 3\mathchar`-sym}=6$ for fundamental, adjoint, rank-2 symmetric, and rank-3 symmetric representations, respectively. 
Casimir scaling means that string tension associated with representation r is proportional to $C_{\rm r}$. Specifically, the Polyakov loop correlator is
\begin{align}
\left\langle
\chi_{\rm r}(P_{\vec{x}})
\cdot
\chi_{\rm r}(P_{\vec{x}+L\hat{u}})
\right\rangle
\sim
e^{-\beta C_{\rm r}L\sigma_0^2}\, ,
\label{eq:2-pt-fnc-and-string-tension}
\end{align}
with $\sigma_0$ independent of $r$ in the regime where Casimir scaling holds.
This scaling does not hold at very long distances where strings can break due to pair creation of particles such as quarks or gluons.   
\subsubsection*{Random walk and string tension}
Let us see how the computation of string tension is related to random walk. 
Character $\chi_{\rm r}$ associated with irreducible representation r is defined by taking the trace of the representation matrix.
Specifically, 
\begin{align}
\chi_{\rm r}(P_{\vec{x}})
=
\mathrm{Tr}_{\rm r}\left( R^{\rm (r)}(P_{\vec{x}})\right)
\end{align}
is the Polyakov loop in representation r at point $\vec{x}$. 
Because of
\begin{align}
\chi_{\rm r}(P_{\vec{x}})
=
\chi_{\rm r}(P'_{\vec{x}})
\end{align}
and 
\begin{align}
\chi_{\rm r'}(P_{\vec{x}+L\hat{u}})
=
\chi_{\rm r'}(P'_{\vec{x}+L\hat{u}})
=
\chi_{\rm r'}
\left(
P_{\vec{x}}^{\prime }
\cdot
(P_{\vec{x}}^{\prime -1}
\cdot
P'_{\vec{x}+L\hat{u}})
\right)\, , 
\end{align}
the two-point function of Polyakov loops can be written as
\begin{align}
\left\langle
\left(
\chi_{\rm r}(P_{\vec{x}})
\right)^\ast
\cdot
\chi_{\rm r'}(P_{\vec{x}+L\hat{u}})
\right\rangle
=
\left\langle
\left(
\chi_{\rm r}(P'_{\vec{x}})
\right)^\ast
\cdot
\chi_{\rm r'}
\left(
P_{\vec{x}}^{\prime}
\cdot
(P_{\vec{x}}^{\prime -1}
\cdot
P'_{\vec{x}+L\hat{u}})
\right)
\right\rangle\, . 
\end{align}

So far, $\langle\ \rangle$ meant the expectation value defined by path integral. From here on, we interpret $\langle\ \rangle$ as the average over slowly varying Haar random configurations $\{P'_{\vec{x}}\}$, neglecting corrections to slowly varying Haar randomness. (As we will see, this approximation fails if $L$ is too large.)

We consider random walk from $P'_{\vec{x}}$ to $P'_{\vec{x}+L\hat{u}}$. 
The starting point $P'_{\vec{x}}$ is Haar random, and it is not correlated to the displacement $P_{\vec{x}}^{\prime -1}P'_{\vec{x}+L\hat{u}}$. 
Therefore, we can treat $P'_{\vec{x}}$ and $P_{\vec{x}}^{\prime -1}P'_{\vec{x}+L\hat{u}}$ as independent variables and integrate out $P'_{\vec{x}}$ by using the orthogonality condition of representations under the Haar-random average, 
\begin{align}
\frac{1}{{\rm Vol}(\mathrm{SU}(N))}
\int dP
(R^{({\rm r})}_{ij}(P))^\ast
\cdot
R^{({\rm r'})}_{kl}(P)
=
d_{\rm r}^{-1}
\delta_{\rm rr'}\delta_{il}\delta_{kj}\, 
\end{align}
to obtain
\begin{align}
\langle (\chi_{\rm r}(P_{\vec{x}}))^\ast
\cdot
\chi_{\rm r'}(P_{\vec{x}+L\hat{u}})\rangle
&=
d_{\rm r}^{-1}\delta_{rr'}
\left\langle \chi_{\rm r}(P_{\vec{x}}^{\prime -1}
\cdot
P'_{\vec{x}+L\hat{u}})\right\rangle\, , 
\label{sec:getting_matrix_product}
\end{align}
and hence, by using \eqref{eq:2-pt-fnc-and-string-tension}, we get 
\begin{align}
d_{\rm r}^{-1}\delta_{rr'}
\left\langle \chi_{\rm r}(P_{\vec{x}}^{\prime -1}
\cdot
P'_{\vec{x}+L\hat{u}})\right\rangle
\sim
e^{-\beta C_{\rm r}L\sigma_0^2}\, .
\end{align}
Combined with \eqref{eq:path-ordering} and \eqref{eq:path-ordering-r}, the computation of the two-point function reduces to a sort of random walk on group manifold. The difference from the vanilla random walk problem is that velocity $v$ changes slowly.
\subsubsection*{Organization of this paper}
In the rest of the paper, we will show that confinement with Casimir scaling follows naturally from such a random walk on a group manifold. 
Our argument is based only on the slowly varying Haar randomness, and the choice of gauge group or matter content is irrelevant.
However, corrections to Haar randomness depend on these specifications of the theory, and they are relevant for long-distance physics where the Casimir scaling is lost. 

This paper is organized as follows. Sec.~~\ref{sec:Hanada-Watanabe-review} explains how slowly varying Haar randomness of the Polyakov loop emerges from a generic mechanism of confinement.
Sec.~\ref{sec:Random_matrix_product}, the main part of this paper, derives Casimir scaling from a random walk. The simplest, `vanilla' random walk is studied in Sec.~\ref{sec:vanilla}. It is solvable, and Casimir scaling can be analytically derived. A more sophisticated variant of random walk is studied in Sec.~\ref{sec:slowly varying-random-walk}. We find a neat correspondence with a lattice regularization and confirm Casimir scaling and consistency with the microscopic mechanism of confinement based on the similarity to Bose-Einstein condensation~\cite{Hanada:2023rlk,Hanada:2023krw,Hanada:2020uvt}. 
Note, however, that we only \textit{model} the random walk and details that can affect the short-distance behavior might depend on the specific model. We expect that different confining theories can show different short-distance behaviors because of the dependence on such details. Larger distance behavior, on the other hand, emerges quite universally from the different models of the random walk.
In Sec.~\ref{sec:string-breaking}, it is discussed that corrections to exact Haar randomness lead to a breaking of Casimir scaling. Sec.~\ref{sec:discussion} is devoted to discussion. 
\section{Microscopic origin of slowly varying Haar randomness}\label{sec:Hanada-Watanabe-review}
\hspace{0.51cm}
In this section, we review the origin of the slowly varying Haar randomness of the Polyakov line~\cite{Hanada:2023rlk}.
As a first step, we rewrite the partition function in terms of an extended Hilbert space ($\mathcal{H}_{\text{ext}}$), including also non-gauge invariant states, and a projection to the gauge-invariant part ($\mathcal{H}_{\text{inv}}$) by integration over all gauge variations (gauge orbit).
For a gauge  theory with gauge group $G$, the canonical partition function can be written as 
\begin{eqnarray}
    Z(T) = \mathrm{Tr}_{\mathcal{H}_{\text{inv}}} e^{-\hat{H}/T} = \frac{1}{\text{vol}G}\int_G dg \mathrm{Tr}_{\mathcal{H}_{\text{ext}}}\left(\hat{g} e^{-\hat{H}/T} \right)\; .
    \label{eq:partition_function}
\end{eqnarray}
where the integral is taken over the Haar measure. A simple but crucial fact is that $\hat{g}$ in the expression above corresponds to the Polyakov line~\cite{Hanada:2020uvt}.
\subsection{$\mathrm{S}_N$ invariance and Bose-Einstein condensation}
\hspace{0.51cm}
To understand the origin of slowly varying Haar randomness, it is instructive to review the connection between the Bose-Einstein Condensation (BEC) and confinement in gauge theories~\cite{Hanada:2023rlk,Hanada:2020uvt}. In this case the gauge group is the permutations $\mathrm{S}_N$ of $N$ particles, the group integral becomes a sum over all possible group elements.

Consider a system of $N$ indistinguishable bosons with a harmonic potential. The Hamiltonian is given by 
\begin{eqnarray}
    \hat{H} = \frac{1}{2} \sum_{i=1}^N \left(\hat{p}^2_{xi} + \hat{p}^2_{yi} + \hat{p}^2_{zi} + \hat{x}^2_i + \hat{y}^2_i +\hat{z}^2_i\right)\, . 
\end{eqnarray}
This system has a gauged $\mathrm{S}_N$ permutation group symmetry. Namely, states connected by $\mathrm{S}_N$ transformation (exchange of labels $1,2,\cdots,N$) must be identified.
A generic state spanning the extended Hilbert space $\mathcal{H}_{\text{ext}}$ is given in the Fock basis as 
\begin{eqnarray}
    |\vec{n}_1,\vec{n}_2, \ldots, \vec{n}_N \rangle = \prod_{i=x,y,z} \frac{a^{\dag n_{1i}}_{1i}}{\sqrt{n_{1i}!}} \frac{a^{\dag n_{2i}}_{2i}}{\sqrt{n_{2i}!}} \ldots \frac{a^{\dag n_{Ni}}_{Ni}}{\sqrt{n_{Ni}!}} | \textbf{0} \rangle
\end{eqnarray}
where $\vec{n}_i = (n_{ix},n_{iy},n_{iz})$ denotes the occupation numbers for each boson. The Fock vacuum is described as $|\textbf{0}\rangle = |0,0,\ldots,0 \rangle$.

For the system of $N$ bosons under consideration here with a gauged $S_N$ symmetry, the partition function takes the form,
\begin{eqnarray}
    Z(T) &=& \frac{1}{N!} \sum_{\sigma \in S_N}  \mathrm{Tr}_{\mathcal{H}_{\text{ext}}} \left( \hat{\sigma} e^{-\hat{H}/T} \right) \nonumber \\
    &=& \frac{1}{N!} \sum_{\sigma \in S_N} \sum_{\vec{n}_N,\vec{n}_N,\ldots,\vec{n}_N} \langle \vec{n}_1,\vec{n}_2,\ldots,\vec{n}_N| \hat{\sigma} e^{-\hat{H}/T} |\vec{n}_1,\vec{n}_2,\ldots,\vec{n}_N \rangle \nonumber \\
    &=& \frac{1}{N!} \sum_{\vec{n}_1,\vec{n}_2,\ldots,\vec{n}_N} e^{-(E_{\vec{n}_1}+\ldots+E_{\vec{n}_N})/T} \sum_{\sigma \in S_N} \langle \vec{n}_1,\vec{n}_2,\ldots,\vec{n}_N | \vec{n}_{\sigma(1)},\vec{n}_{\sigma(2)},\ldots,\vec{n}_{\sigma(N)} \rangle \nonumber \\
    &=& \frac{1}{N!} \sum_{\vec{n_1},\vec{n_2},\ldots,\vec{n_N}} e^{-(E_{\vec{n}_1}+\ldots+E_{\vec{n}_N})/T} \sum_{\sigma \in S_N} \delta_{\vec{n}_1 \vec{n}_{\sigma(1)}} \delta_{\vec{n}_2 \vec{n}_{\sigma(2)}}\ldots \delta_{\vec{n}_N \vec{n}_{\sigma(N)}} \nonumber \\
    &=& \frac{1}{N!} \sum_{\vec{n_1},\vec{n_2},\ldots,\vec{n_N}} e^{-(E_{\vec{n}_1}+\ldots+E_{\vec{n}_N})/T} V_{|\vec{n}_1,\vec{n}_2,\ldots,\vec{n}_N \rangle}
\end{eqnarray}
Here, $V_{|\vec{n}_1,\vec{n}_2,\ldots,\vec{n}_N \rangle} \equiv \sum_{\sigma \in S_N} \delta_{\vec{n}_1 \vec{n}_{\sigma(1)}} \delta_{\vec{n}_2 \vec{n}_{\sigma(2)}}\ldots \delta_{\vec{n}_N \vec{n}_{\sigma(N)}} $ is the volume of the stabilizer subgroup of the state $|\vec{n}_1,\vec{n}_2,\ldots,\vec{n}_N \rangle$ i.e., the number of the elements of $S_N$ that leave the state invariant. If all the particles have different occupation numbers ($\vec{n}_i \neq \vec{n}_j$ for all $i \neq j$), then the stabilizer of such a state is just the identity element $\sigma = \mathbf{1}$. The contribution of such a state to the partition function thus remains suppressed by a factor of $1/N!$. However, if all the particles have the same occupation number ($\vec{n}_1 = \ldots = \vec{n}_N = \vec{n}$), then the stabilizer of such a state is the entire permutation group $S_N$ with $V_{|\vec{n},\ldots,\vec{n} \rangle} = N!$. The suppression factor in the partition function thus experiences an enhancement and becomes equal to 1 and such a state is then favoured. This is exactly what happens in the ground state $|\textbf{0}\rangle$ ($\vec{n}_1 = \ldots = \vec{n}_N = \vec{0}$). More generally, if the energy is distributed to a part of particles and the rest remains in the one-particle ground state (say, $\vec{n}_1,\cdots,\vec{n}_M$ becomes nonzero while $n_{M+1}=\cdots\vec{N}=\vec{0}$), a large enhancement factor $(N-M)!$ is obtained. In general, the optimal value of $M$ between $0$ and $N$ is realized so that the combination of such an enhancement factor and the entropy of the excited sector is maximized. This is the BEC of $N-M$ particles. This mechanism works with finite interactions as well, and unbroken $\mathrm{S}_{N-M}$ symmetry in the extended Hilbert space characterizes BEC. For $M=0$, all `Polyakov line' $\sigma\in\mathrm{S}_N$ contribute equally, which means the Haar-randomness. 
\subsection{$\mathrm{SU}(N)$ invariance and confinement}
\begin{figure}[htbp]
\begin{center}
\scalebox{0.25}{
\includegraphics{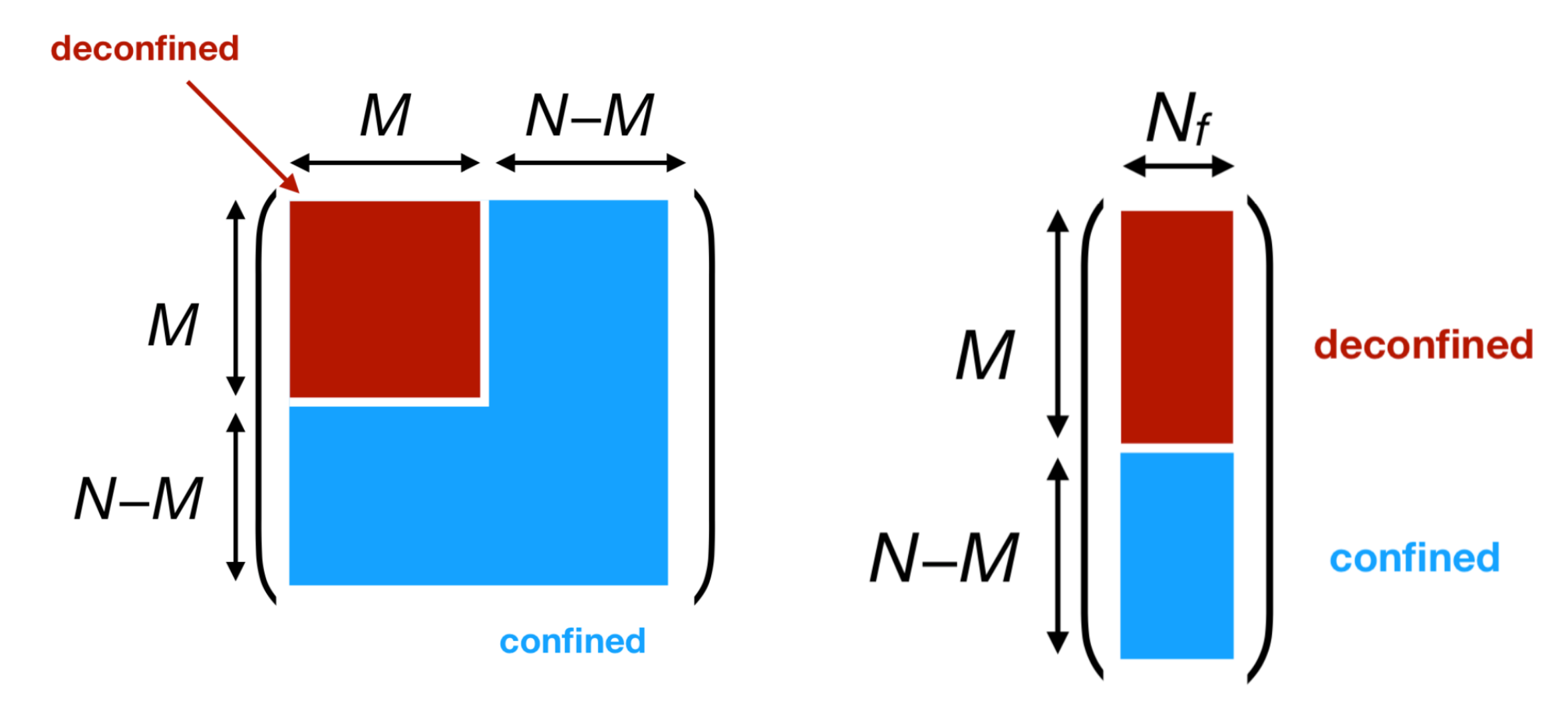}}
\end{center}
\caption{
Partial deconfinement in SU($N$) QCD with $N_f$ fundamental quarks. [Left] Gauge field. $M\times M$ sub-matrix is deconfined.
[Right] Quarks. $M$ components are deconfined. Any embeddings of SU($M$) into SU($N$) are equivalent because of gauge symmetry. This figure is taken from Ref.~\cite{Hanada:2019czd}.  
}\label{fig:partial_deconfinement}
\end{figure}

\hspace{0.51cm}
Coming back from BEC to our original question about gauge theories including $\mathrm{SU}(N)$ pure Yang-Mills theory, a similar enhancement can be observed. In general, an $\mathrm{SU}(M)$ subgroup can be deconfined, while the rest of the degrees of freedom remain confined~\cite{Hanada:2016pwv,Berenstein:2018lrm,Hanada:2018zxn,Hanada:2019czd} as illustrated in Fig.~\ref{fig:partial_deconfinement}. Roughly speaking, such \textit{partial deconfinement} is characterized by an unbroken $\mathrm{SU}(N-M)$ symmetry. Since this is considered in the extended Hilbert space, there is no conflict with gauge invariance; simply, all embeddings of SU($M$) into SU($N$) are equivalent. 
The completely confined phase we are discussing in this paper is characterized by unbroken $\mathrm{SU}(N)$ symmetry in the extended Hilbert space. Let us elaborate on this point for the case of Yang-Mills theory and see how the slowly varying Haar randomness of the Polyakov line emerges. The differences from the example of BEC are that the gauge group is bigger (SU($N$) instead of S$_N$) and that we need to deal with spatial coordinates because we deal with QFT rather than quantum mechanics.

For concreteness, we consider pure Yang-Mills theory and use a lattice regularization with lattice spacing $a$ and unitary link variables $U_{\vec{n},\mu}$ which is related to the Hermitian gauge field $A_\mu$ as $U_{\vec{n},\mu}=e^{iag_{\rm YM}A_{\vec{n},\mu}}$. Here, $\vec{n}$ is an integer-valued vector that specifies a spatial site on the lattice, and $\mu$ is a spatial direction $x,y,\ldots$.
In the extended-Hilbert space picture, wave functions are defined on $\prod_{\mu,\vec{n}}[\mathrm{SU}(N)]_{\mu,\vec{n}}$, where $[\mathrm{SU}(N)]_{\mu,\vec{n}}$ corresponds to each link. Specifically, on each link, we consider a linear combination of coordinate eigenstates $|U_{\vec{n},\mu}\rangle$. The confined vacuum in or sufficiently close to the continuum limit is a wave packet localized around $U_{\vec{n},\mu} = \mathbf{1}$ up to gauge transformation. (As we will see shortly, a large enhancement factor is associated with such a state, which justifies its dominance in the partition function.)
Wave packets connected by a (spacial) gauge transformation $\Omega \in SU(N)$ are equivalent. Therefore, we should consider wave packets localized around $U_{\vec{n},\mu}=\Omega_{\vec{n}}^{-1}\Omega_{\vec{n}+\hat{\mu}}$.

\begin{figure}[htbp]
\begin{center}
\scalebox{0.3}{
\includegraphics{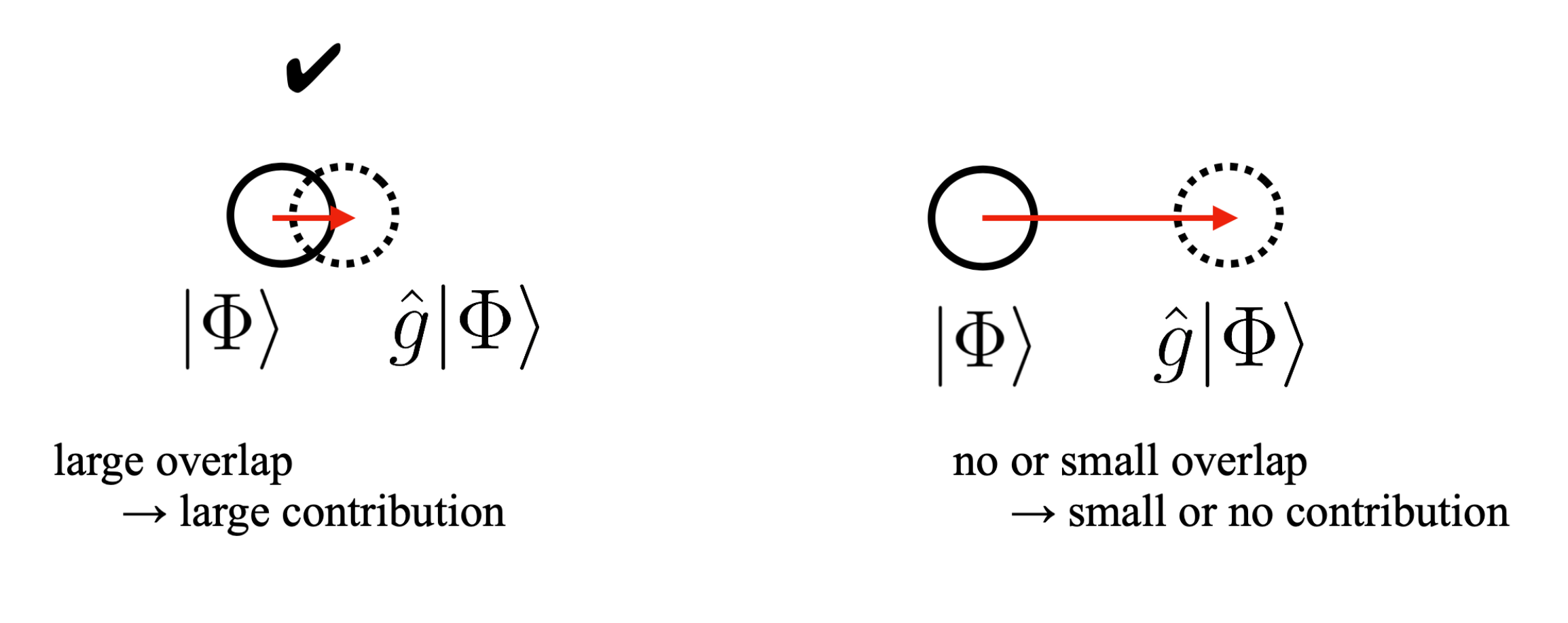}}
\end{center}
\caption{
For a wave packet $|\Phi\rangle$, such $g$ that does not move $|\Phi\rangle$ too much can lead to an enhancement factor and give a large contribution to the partition function.
}\label{fig:overlap}
\end{figure}

Corresponding to
$\langle \vec{n}_1,\vec{n}_2,\ldots,\vec{n}_N | \vec{n}_{\sigma(1)},\vec{n}_{\sigma(2)},\ldots,\vec{n}_{\sigma(N)} \rangle$, we have a factor $\langle\Phi|\hat{g}|\Phi\rangle$ in the partition function, where we can take $|\Phi\rangle$ to be a wave packet. By definition, this factor measures overlap between $|\Phi\rangle$ and $\hat{g}|\Phi\rangle$; see Fig.~\ref{fig:overlap}. Because the Polyakov lines correspond to $\hat{g}$, such Polyakov lines that do not move the wave packet too much can contribute to the enhancement factor. 
Let us consider the action of the Polyakov loop corresponding to `global' $SU(N)$ transformation, $P_{\vec{n}} \equiv \Omega_{\vec{n}}^{-1} V \Omega_{\vec{n}}$.  It is easy to see that the wave packet under consideration does not move regardless of the choice of $V$, because 
\begin{eqnarray}
    P_{\vec{n}}^{-1} (\Omega_{\vec{n}}^{-1} \Omega_{\vec{n}+\hat{\mu}}) P_{\vec{n}+\hat{\mu}} = \Omega_{\vec{n}}^{-1} V^{-1} \Omega_{\vec{n}} (\Omega_{\vec{n}}^{-1} \Omega_{\vec{n}+\hat{\mu}}) \Omega_{\vec{n}+ \hat{\mu}}^{-1} V \Omega_{\vec{n}+\hat{\mu}} = \Omega_{\vec{n}}^{-1} \Omega_{\vec{n}+\hat{\mu}}\, . 
\end{eqnarray}
Thus, the vacuum is invariant under this `global' $SU(N)$ transformation, leading to an enhancement factor as in the BEC case. The enhancement factor scales with $N$ as $e^{N^2}$.
For partially-deconfined states shown in Fig.~\ref{fig:partial_deconfinement}, 
\begin{align}
V = 
\left(
\begin{array}{cc}
\textbf{1}_{M} & \textbf{0}\\
\textbf{0} & \tilde{V}
\end{array}
\right)
\end{align}
with any $\tilde{V}\in\textrm{SU}(N-M)$ could leave the states invariant. This leads to an enhancement factor $\sim e^{(N-M)^2}$. 

Let us elaborate on the argument above and show that the enhancement factor can increase exponentially with volume. Roughly speaking, we just promote a `global' transformation to `slowly varying' transformation.
Let us start with the case of the confined vacuum ($M=0$). We consider a `local' $SU(N)$ transformation with $V$ now being dependent on the site $\vec{n}$ with the Polyakov line taking the form $P_{\vec{n}} \equiv \Omega_{\vec{n}}^{-1} V_{\vec{n}} \Omega_{\vec{n}}$:
\begin{eqnarray}
    P_{\vec{n}}^{-1} U_{\vec{n},\mu} P_{\vec{n}+\hat{\mu}} = \Omega_{\vec{n}}^{-1} V_{\vec{n}}^{-1} \Omega_{\vec{n}} (\Omega_{\vec{n}}^{-1} \Omega_{\vec{n}+\hat{\mu}}) \Omega_{\vec{n}+ \hat{\mu}}^{-1} V_{\vec{n}+\hat{\mu}} \Omega_{\vec{n}+\hat{\mu}} = \Omega_{\vec{n}}^{-1} V_{\vec{n}}^{-1} V_{\vec{n}+\hat{\mu}} \Omega_{\vec{n}+\hat{\mu}}
\end{eqnarray}
If $V_{\vec{n}}$ is slowly varying, i.e., $V_{\vec{n}}^{-1} V_{\vec{n}+\hat{\mu}}$ is close to $\mathbf{1}$, and we can still have a significant enhancement factor as before. For the confined vacuum, we can choose $V_{\vec{n}}$ arbitrarily as long as it is slowly varying, indeed leading to a large enhancement factor that increases with spatial volume. 
Intuitively, we can split the volume into multiple pieces that are large enough so that the correlation between Polyakov lines in different pieces is negligible, and then, we can associate $\sim e^{N^2}$ to each piece, which leads to the scaling of the enhancement factor $e^{N^2\times\mathrm{volume}}$. 

\subsection{Continuum limit and smearing}\label{sec:renormalization}
\hspace{0.51cm}
In the previous subsection, we explained how the bare Polyakov line and symmetry of wave function on each spatial link are related. When we take continuum limit ($a\to 0$), we need to see how renormalization affects the logic. 

In the continuum limit with fixed temperature, the expectation value of the bare Polyakov loop vanishes even in the deconfined phase for any representation. A multiplicative renormalization factor is needed to keep the expectation value finite~\cite{Gupta:2007ax,Gupta:2006qm,Mykkanen:2012ri}. This means that Polyakov line becomes Haar random even in the deconfined phase. In terms of symmetries, this means that it is hard to see the difference between the ground state and  excited states as long as we focus on the wave function on each link. Note also that, associated with ultraviolet divergence, the size of the wave packet can grow as $a\to 0$ for 3d and 4d theories,\footnote{
For Euclidean lattice, the expectation value of plaquette scales as $e^{-a^2}$ and $e^{-a}$ for 2d and 3d, respectively. 
Before taking expectation value, plaquette is $U_{\mu,\vec{x}}U_{\nu,\vec{x}+\hat{\mu}}U^\dagger_{\mu,\vec{x}+\hat{\nu}}U^\dagger_{\nu,\vec{x}}\simeq e^{ia^2F_{\mu\nu,\vec{x}}}\sim e^{ia}$ for 2d and $\sim e^{i\sqrt{a}}$ for 3d. Because $U_\mu\simeq e^{iaA_{\mu,\vec{x}}}$, typical value of $A_{\mu,\vec{x}}$ is order 1 for 2d and order $a^{-1/2}$ for 3d. 
} which may destroy the slowly varyingness of Polyakov line. Still, however, infrared properties can be significantly different because the number of links increases and small change at each link can accumulate. 

To capture the symmetries and properties relevant for thermodynamics, we should not use bare Polyakov line with too small lattice spacing. 
To renormalize the Polyakov line, we can combine the facts that the Polyakov line can be written as a sum of characters and each character can be renormalized in a standard manner~\cite{Gupta:2007ax,Gupta:2006qm,Mykkanen:2012ri}. 
We could also use a smeared gauge field and define a smeared version of the Polyakov loop.   

As a concrete example, let us consider the smearing via gradient flow~\cite{Luscher:2010iy,Datta:2015bzm,Petreczky:2015yta}. 
In this smearing scheme, smeared gauge field is defined at each link on the original lattice. By replacing the original gauge field with smeared one, the smeared Polyakov line is obtained.  
Because we are interested in correlators at not-so-short distance and at finite-temperature (specifically, in the confined phase), we should take smearing length at most order of $\Lambda_{\rm QCD}^{-1}$. Then, the smearing does not affect the long-distance behavior and hence the string tension. The smearing makes the spatial-coordinate dependence of $P'_{\vec{x}}$ mild and introduces the slowly varying Haar randomness starting at short distance; see also Sec.~\ref{sec:epsilon-estimate}. We cannot use such a Polyakov line made of smeared field to study short-distance physics below the smearing length, but that is not a problem for us as long as we discuss Casimir scaling at sufficiently long distance.\footnote{
More precisely, the connected part of two-point function is not affected.}
What we will claim is the existence of a smearing scheme consistent with slowly varying Haar randomness is a sufficient condition for the Casimir scaling to be valid.

\subsection{Slowly varying Haar randomness}\label{sec:epsilon-estimate}
\hspace{0.51cm}
So far, we have seen that a large enhancement factor is indeed associated with the confined vacuum. We have also seen that Polyakov lines contributing to the enhancement factor are written as $P_{\vec{n}} \equiv \Omega_{\vec{n}}^{-1} V_{\vec{n}} \Omega_{\vec{n}}$, where $V_{\vec{n}}$ can be arbitrary as long as it is slowly varying. Typical Polyakov lines dominating the partition function (equivalently, typical Polyakov lines obtained from dominant configurations in the Euclidean path integral) are the ones that contribute to the enhancement factor. Therefore, typical Polyakov lines are slowly varying SU($N$) Haar random up to gauge transformation~\cite{Hanada:2023rlk}. For the same logic, the Polyakov line restricted to the confined sector of the partially-deconfined phase should be slowly varying SU($N-M$) Haar random up to gauge transformation.

`Slowly varying' is admittedly a vague expression. Below, we give an order estimate for the rate of change. 
We want to know how slowly a typical Polyakov line changes, assuming the confined vacuum represented by a wave packet localized around a pure-gauge configuration. Because a typical Polyakov line must lead to a large enhancement factor, $V_{\vec{n}}^{-1} V_{\vec{n}+\hat{\mu}}$ has to be sufficiently close to the identity so that the ground-state wave function does not move too much (Fig.~\ref{fig:overlap}). Specifically, if $V_{\vec{n}}^{-1} V_{\vec{n}+\hat{\mu}}$ is written as $V_{\vec{n}}^{-1} V_{\vec{n}+\hat{\mu}}\sim e^{i\ell\tilde{X}}$ where $\tilde{X}$'s entries are of order $1$, then $\ell$ must be smaller than the radius of the wave packet. 

Let us start with the two-dimensional Yang-Mills theory, which does not require smearing.
Although it is straightforward to solve it in the path-integral formalism, let us use the Hamiltonian formulation to derive the same result. (The same result is derived in Appendix.~\ref{appendix:2dYM} by using path integral.) The Hamiltonian contains only the electric term, $\hat{H} =\frac{a}{2}\mathrm{Tr}\sum_{\vec{x}} \hat{E}_{\vec{x}}^2$, and there is no interaction between links. This is essentially $({\rm momentum})^2$ on the group manifold. As temperature goes up, higher momentum contributes more, or equivalently, typical wave packet shrinks. The energy is proportional to $T$, and hence, typical momentum is $\sqrt{\frac{T}{a}}=(\beta a)^{-1/2}$, and the radius of typical wave packet scales as $\sqrt{\beta a}$. Therefore, we expect $V_{n}^{-1}V_{n+1} \sim e^{i\sqrt{\beta a}\tilde{X}}$. Due to the absence of the interaction in the Hamiltonian, there is no correlation between $V_{n}^{-1}V_{n+1}$ for different $n$. In other words, `velocity' is totally random at each link. Therefore, $V_{m}^{-1}V_{n}\sim e^{i\sqrt{\beta a|m-n|}\tilde{X}}$. 

Next, let us consider 3d and 4d Yang-Mills theory, which is more complicated because the bare Polyakov line can become Haar random in the continuum limit even in the deconfined phase. To take the continuum limit, we consider the loops made of smeared gauge field. Because of the randomness of the bare loop, smeared loop should also exhibit the random-walk nature. At a smearing radius of order $\Lambda_{\rm QCD}^{-1}$, the natural value of $v$ is of order $a^0$.

Let us also comment on the `acceleration', i.e., the rate of change of $v$. 
In 3d and 4d theories, $v$ in \eqref{eq:path-ordering} has to be smooth because $P_{\vec{x}}^{\prime -1}P'_{\vec{y}}$ depends only on the endpoints $\vec{x}$ and $\vec{y}$ and does not change depending on the path connecting these two points. For this reason, the Polyakov lines exhibit random walks with slowly changing velocity, which is different from the most common random walk in which the velocity at each step changes abruptly.
In 2d theories, $v$ does not have to be continuous and the Polyakov line can literally random walk. This has an important consequence as we will discuss in Sec.~\ref{sec:(1+1)-d}.
\section{Casimir scaling from random walk}\label{sec:Random_matrix_product}
\begin{figure}[htbp]
\begin{center}
\scalebox{0.25}{
\includegraphics{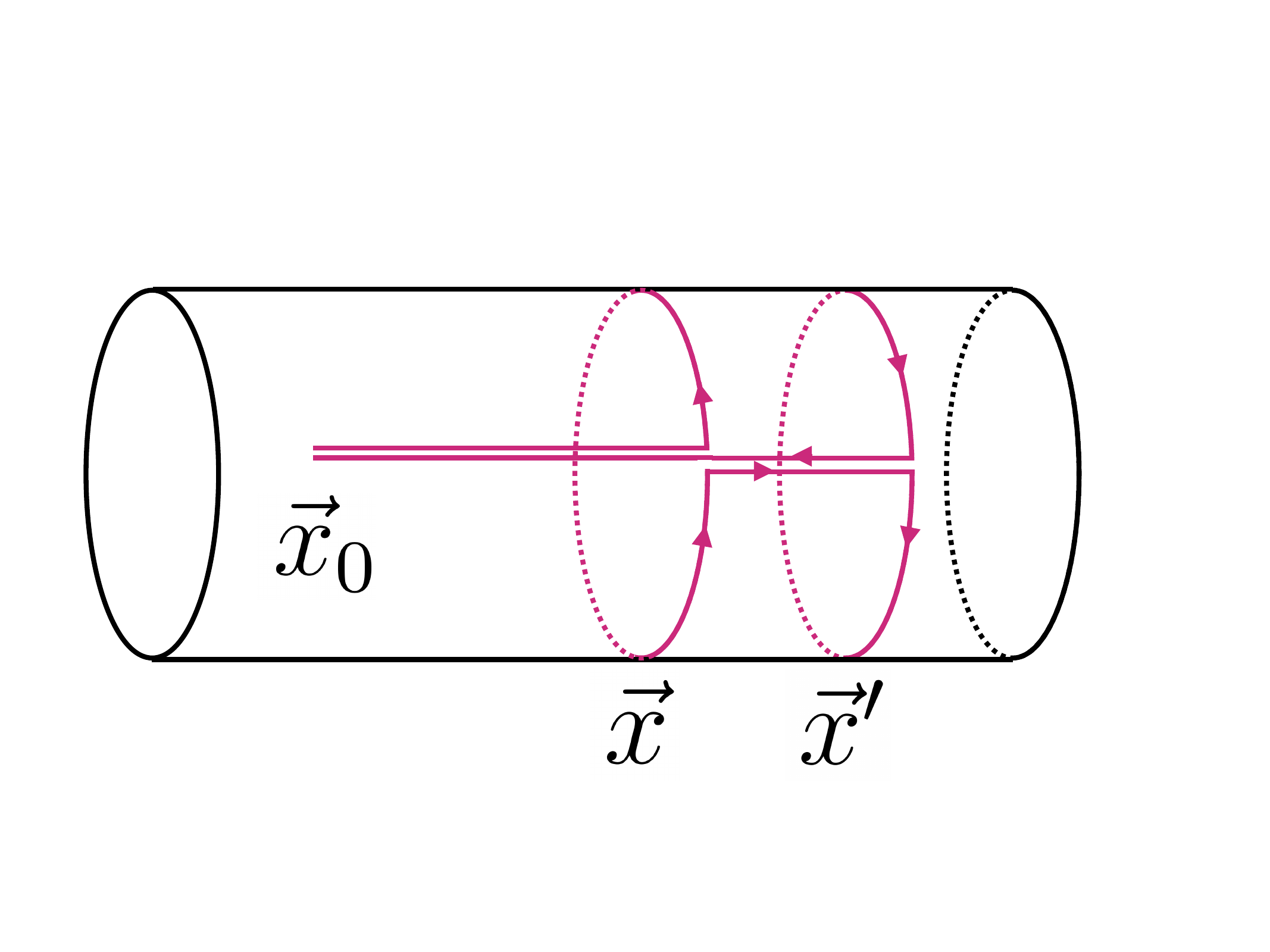}}
\end{center}
\caption{Contour corresponding to
$P_{\vec{x}}^{\prime -1}
P'_{\vec{x}'}$
to illustrate the definition of $W_j$ in \eqref{eq:Wj}.
}\label{fig:Figure_W}
\end{figure}

\hspace{0.51cm}
In this section, we assume the exact Haar randomness and derive the linear confinement potential with Casimir scaling. 
Hence, our task is to estimate $
\left\langle \chi_{\rm r}(P_{\vec{x}}^{\prime -1}
P'_{\vec{x}+L\hat{u}})\right\rangle$ assuming the random walk \eqref{eq:path-ordering}.
To evaluate \eqref{eq:path-ordering}, let us introduce step size $a$, where $Ka=L$, and take a limit of $a\to 0$. 
(We use $a$ because it is analogous to the lattice spacing in lattice regularized theories.)
We define $W_j$ ($j=1,2,\cdots,K$) by 
\begin{align}
W_j
\equiv
(P'_{\vec{x}+(j-1)a\hat{u}})^{-1}
P'_{\vec{x}+ja\hat{u}}\, . \label{eq:Wj}
\end{align}
These can be combined to
\begin{align}
P_{\vec{x}}^{\prime -1}
P'_{\vec{x}+L\vec{u}}
=
W_1W_2\cdots W_K\, , 
\end{align}
as illustrated in Fig.~\ref{fig:Figure_W}. 

Below, let us first consider the vanilla random walk as a solvable warm-up exercise, and then consider a more realistic setup that takes into account the correlation between $W_i$'s. 
\subsection{Vanilla random walk}\label{sec:vanilla}
\hspace{0.51cm}
As a crude approximation, let us treat $W_1,\cdots,W_K$ totally random, even neglecting the correlation of $W_i$ and $W_{i+1}$. 
This is the vanilla random walk. 
Apparently, we cannot always take the limit of $a\to 0$ in this case. To keep this in mind, let us use $s$ instead of $a$. 
Physically, $s$ is a length scale needed for $v$ in \eqref{eq:path-ordering} to change significantly and the correlation between $v(L)$ and $v(L+s)$ becomes small.  
With this approximation, the average of a product becomes the product of the average:
\begin{align}
\left\langle W^{\rm (r)}_1W^{\rm (r)}_2\cdots W^{\rm (r)}_K\right\rangle
=
\left\langle W^{\rm (r)}_1\right\rangle
\left\langle W^{\rm (r)}_2\right\rangle
\cdots 
\left\langle W^{\rm (r)}_K\right\rangle
=
\left(
\left\langle W^{\rm (r)}\right\rangle
\right)^K\, .
\end{align}
This expression already tells us an exponential decay. 
To see the Casimir scaling, we need a little bit more math. 
We assume $W=e^{i\Delta X}$, where $X$ is random Gaussian with variance 1 and $\Delta$ is a small number. Then, 
\begin{align}
\left\langle W^{\rm (r)}\right\rangle
&=
\left\langle \textbf{1} + i\Delta x^\alpha T^{\rm (r)}_\alpha - \frac{\Delta^2 x^\alpha x^\beta}{2}T^{\rm (r)}_\alpha T^{\rm (r)}_\beta+\cdots\right\rangle
\nonumber\\
&=
 \textbf{1} - \frac{\Delta^2}{2}(T^{\rm (r)}_\alpha)^2+\cdots
 \nonumber\\
&=
 \textbf{1} - 2\Delta^2 C_{\rm r} \textbf{1}+\cdots\, . 
\end{align}
Hence, for small $\Delta$,
\begin{align}
\left\langle W^{\rm (r)}\right\rangle
\simeq
e^{-2\Delta^2C_{\rm r}} \textbf{1} 
\end{align}
and hence
\begin{align}
\left\langle W^{\rm (r)}_1\right\rangle
\left\langle W^{\rm (r)}_2\right\rangle
\cdots 
\left\langle W^{\rm (r)}_K\right\rangle
\simeq
e^{-2\Delta^2C_{\rm r}K} \textbf{1} \, .
\end{align}
Therefore, 
\begin{align}
\langle (\chi_{\rm r}(P_{\vec{x}}))^\ast\chi_{\rm r'}(P_{\vec{x}+L\hat{u}})\rangle
\simeq
\delta_{rr'}
e^{-2\Delta^2C_{\rm r}K}
\end{align}
up to $O(\Delta^4)$ terms in the exponent. Therefore, by identifying $\Delta$ with physical parameters as 
\begin{align}
\Delta^2
=
\beta s\sigma_0^2\, , 
\end{align}
string tension for representation r is $C_{\rm r}\sigma_0^2$ if $O(\Delta^4)$ terms are negligible. The dependence on $s$ is natural because after $K$ steps the distance traveled on the group manifold is $\sim\Delta\sqrt{K}\sim \sqrt{L}$, which is a typical case of random walk. 

That terms of order $\Delta^4$ or higher break Casimir scaling would be good news because lattice simulations suggest Casimir scaling is not exact. 

Several references~\cite{Brzoska:2004pi,DiGiacomo:2000irz,Greensite:2006sm} assumed the uncorrelated nature of color flux and explained the Casimir scaling in a way similar to the vanilla random walk in the limit of $\Delta\to 0$. Our discussion above is based on the microscopic mechanism applicable to generic confining gauge theories, and hence, we can give a more precise picture that takes into account a finite amount of correlations, as we will show next. It would be interesting to see if the hypothesis regarding the QCD vacuum proposed in the past can be justified based on our approach.  

As a minor remark, we note that this vanilla random walk works for two-dimensional pure Yang-Mills. 
\subsection{Random walk with gradually changing velocity}\label{sec:slowly varying-random-walk}
\hspace{0.51cm}
In reality, we want the `velocity' $W_i$ to change gradually (except for 2d theory, as we will see in Sec.~\ref{sec:(1+1)-d}). To incorporate this feature, let us consider a model of random walk with $W_i$ defined by 
\begin{align}
W_i
=
Z_{i}\cdots Z_{i+p-1}\, , 
\label{eq:W-as-product}
\end{align}
where $Z_j=e^{i\epsilon X_j}$ and $X_j$ is Gaussian random with a unit variance, 
and take the average of the product of $W$'s as before:
\begin{align}
\langle W_1W_2\cdots W_K\rangle\, . 
\end{align}
Then, $W_i$ changes gradually, and the correlation disappears completely after $p$ steps. Given the decay of the two-point function of Polyakov loops, $ap\sim\frac{1}{\beta\sigma_0^2}$ is a natural identification. (This means $v\sim \beta\sigma_0^2\cdot a^0$, which is consistent with the rough estimate $v\sim a^0$ in Sec.~\ref{sec:epsilon-estimate}.) 
We are interested in the behavior of the correlator as a function of physical distance $L=Ka$.

Unfortunately, we were not able to solve this problem analytically. We performed numerical experiments for SU(2) and SU(3). The results are very clean, and we believe our findings apply to other groups as well. 

\begin{figure}[htbp]
\centering
\scalebox{0.6}{
\input{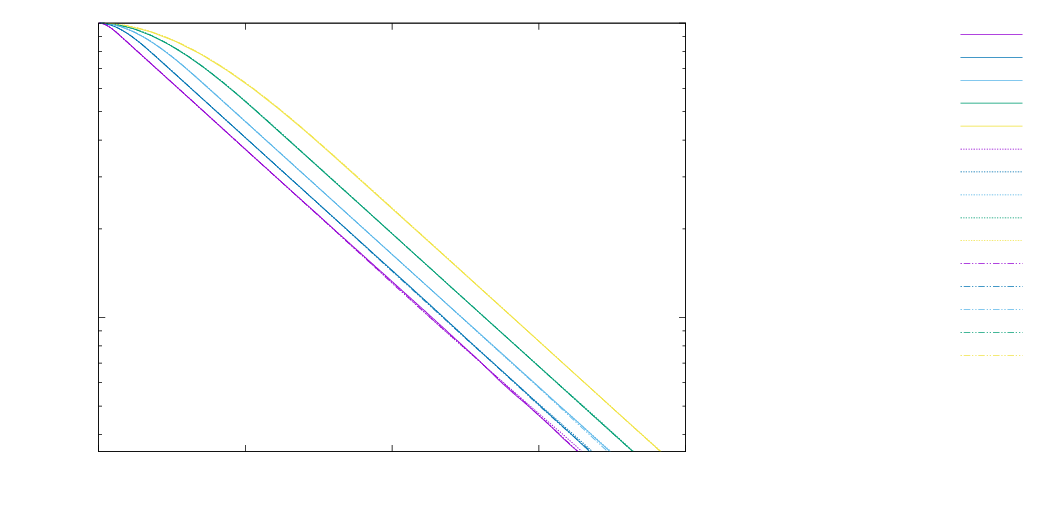}}
\scalebox{0.6}{
\input{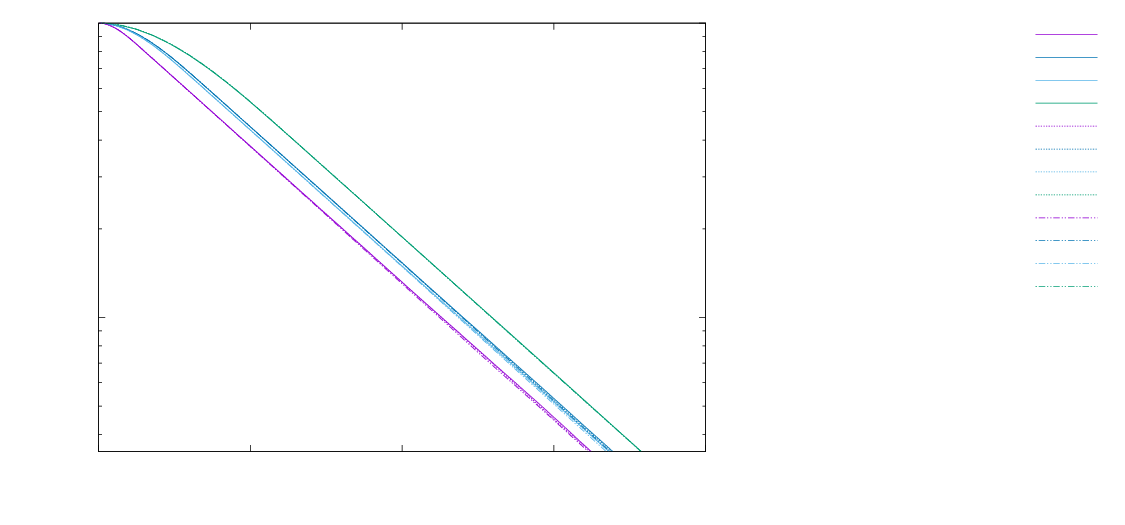}}
\caption{
$\textrm{Tr}\langle W_1...W_K\rangle$ vs $K\cdot (\epsilon p)^2\times C_{\rm r}$ for $p=10, 20, 40$ and $\epsilon^2p^3=\frac{1}{10}$. 
[Top] SU(2), spin $j=\frac{1}{2},1,\frac{3}{2}, 2, \frac{5}{2}$. [Bottom] SU(3), fundamental, rank-2 symmetric, adjoint, rank-3 symmetric. 
We normalized $\textrm{Tr}\langle W_1...W_K\rangle$ by dividing $d_{\rm r}=\textrm{Tr}\textbf{1}$. 
The number of samples for the averaging is 1 million for both SU(2) and SU(3).
}\label{fig:W-various-p}
\end{figure}

As shown in Fig.~\ref{fig:W-various-p}, by taking the horizontal axis to be $K\cdot (\epsilon p)^2\times C_{\rm r}$, we can see that exponential decays with different $p$, $\epsilon$ and different representations line up.  Therefore, by identifying $(\epsilon p)^2$ with $\beta a\sigma_0^2$, we can reproduce $e^{-C_{\rm r}\beta L\sigma_0^2}$. This $\epsilon p$ is a counterpart of $\Delta$ in the vanilla random walk, but now the finite amount of correlation is taken into account. 

Actually, we observe this clean pattern when $\epsilon^2p^3\lesssim 1$. 
With the indentification $(\epsilon p)^2\sim\beta a\sigma_0^2$, this means that $\epsilon^2p^3\sim\beta\sigma_0^2\times ap\lesssim 1$, which is consistent with the identification provided above, i.e., $ap\sim\frac{1}{\beta\sigma_0^2}$.

We emphasize that we only provided a model of the random walk of Polyakov lines satisfying essential features. 
Specifically, $X_j$ may not be exactly Gaussian, or $W_i$ may change in a way different from \eqref{eq:W-as-product}.

\subsection{Simplification for $(1+1)$-dimensional theories}\label{sec:(1+1)-d}
\hspace{0.51cm}
In Sec.~\ref{sec:slowly varying-random-walk}, we have seen that the deviation from vanilla random walk is responsible for the absence of the Casimir scaling at short distance. 
In fact, however, the situation is different for $(1+1)$-dimensional theories.
As commented in Sec.~\ref{sec:epsilon-estimate}, $(1+1)$-dimensional theories are special in that $v$ in \eqref{eq:path-ordering} does not have to be continuous. On a lattice, the vanilla random walk can be justified for any lattice spacing, as long as the corrections to Haar randomness are neglected. Therefore, quite generally, we expect the Casimir scaling also at short distances. 

It is known that the Casimir scaling is exact for $(1+1)$-dimensional pure Yang-Mills theory~\cite{Migdal:1975zg,Witten:1991we,Nguyen:2021naa}. 
With matter fields, infrared behaviors may change drastically. However, we do not expect the short-distance behavior to change, as long as the system remains confined. This provides us with a nontrivial consistency check of the relationship between the Casimir scaling and a random walk on the group manifold. 

\section{String breaking and loss of Casimir scaling}\label{sec:string-breaking}
\hspace{0.51cm}
String breaking is almost trivial when the corrections to Haar-random distribution are taken into account. 
If there is no particular global symmetry that forces the Polyakov loop with representation r to become zero, we expect to have a nonzero expectation value of the one-point function of the form 
\begin{align}
\langle \chi_{\rm r}(P_{\vec{x}})\rangle
=
\langle \chi_{\rm r}(P_{\vec{x}+L\hat{u}})\rangle
\sim
e^{-\beta m_{\rm r}}\, , 
\end{align}
where $m_{\rm r}$ is the mass of the lowest excitation in representation r.
Therefore, we need to take the disconnected part of the two-point function into account: 
\begin{align}
\langle (\chi_{\rm r}(P_{\vec{x}}))^\ast\chi_{\rm r}(P_{\vec{x}+L\hat{u}})\rangle
\sim
e^{-\beta C_{\rm r}L\sigma_0^2}
+
e^{-2\beta m_{\rm r}}\, , 
\end{align}
where the second term is the disconnected part which is dominant at long distances ($C_{\rm r}L\sigma_0^2\gg m_{\rm r}$). 

There must be corrections to the connected part as well. For QCD, the details of the corrections to the connected part may not be important because the disconnected part is nonzero for any representations and dominates long-distance physics. The situation is somewhat different for theories with center symmetry such as SU($N$) pure Yang-Mills. As long as the center symmetry is not broken, the disconnected part vanishes. Furthermore, loops with different center-symmetry charges cannot mix with each other, because the connected part must vanish as well. In the random-walk approach, \eqref{sec:getting_matrix_product} receives some corrections because $P'_{\vec{x}}$ is not exactly Haar random, and we need to consider a random walk on group manifold with some other weight than the Haar measure. We do not find an immediate reason that forbids the mixing in the same center-symmetry-charge sector, and hence, we expect that string tension takes the same value in the same center-symmetry-charge sector regardless of the representation. Precise dependence on charge requires a better understanding and is out of the scope of this paper. See Refs.~\cite{Arcioni:2005iq,Buividovich:2006yj,Buividovich:2007xh} for past attempts to build a random-walk model in the long-distance regime. 
\section{Discussion}\label{sec:discussion}
\hspace{0.51cm}
In the low-temperature regime of confining gauge theories such as QCD, Polyakov lines are slowly varying Haar random modulo exponentially small corrections with respect to the inverse temperature~\cite{Hanada:2023rlk,Hanada:2023krw,Hanada:2020uvt}.
In this paper, we suggested that linear confinement potential with Casimir scaling follows naturally from such a random walk. Strictly speaking, we introduced a model of random walk using a random matrix product. We expect the exponential decay with approximate Casimir scaling to be universal and does not depend much on the details of the random process, while short-distance behaviors depend on details. With exponentially small corrections to Haar randomness, string breaking and loss of Casimir scaling at long distance follow. So, what we have seen is an approximate Casimir scaling in the intermediate distance scale, which is precisely what we expect in confining gauge theories. Our picture can be tested by calculating the Polyakov loops via lattice gauge theory simulation. 

\begin{figure}[htbp]
\begin{center}
\scalebox{0.25}{
\includegraphics{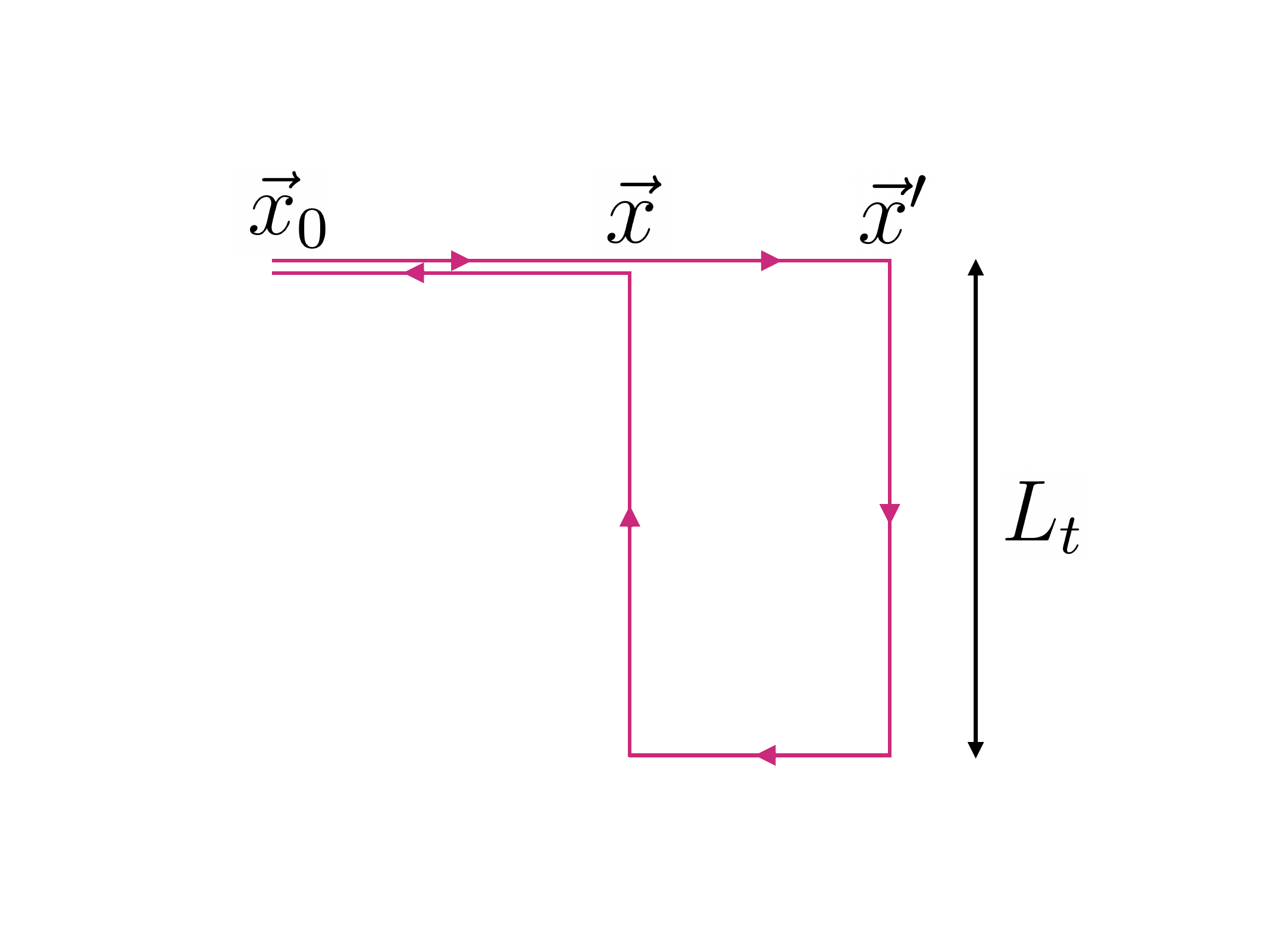}}
\end{center}
\caption{
A Wilson line which can be used to build a rectangular Wilson loop of the size $L_t$ for the time direction and $L$ for the spatial direction. 
}\label{fig:Wilson}
\end{figure}

The mechanism described for the Polyakov loop may generalize to the case of a rectangular Wilson loop of the size $L_t$ for the time direction and $L$ for the spatial direction. Specifically, we could take a product of Wilson lines like in Fig.~\ref{fig:Wilson}. In this case, $(\epsilon p)^{2}\sim L_t a$ would be expected. 
For this scenario to work, we need to take $L_t$ sufficiently large so that the temporal edge inherits the Haar randomness of the Polyakov loop. Note also that, if $L_t$ is too small, the conformal behavior can set in and slowly varying nature can be lost.\footnote{We thank Nadav Drukker for pointing this out for us.}
It would be interesting to use lattice simulation to confirm this mechanism. 

Note that we essentially assumed the mass gap to claim that the correction to Haar randomness is exponentially small. Therefore, we are assuming one of the important features of confinement, and hence, we do not claim we proved confinement. Still, what we observed is a highly nontrivial consequence of the mass gap, and we can find nontrivial connections between good old results and deepen our understanding of confinement. 

More than two decades ago, it was realized that large-$N$ theory in the weak-coupling limit exhibits confinement/deconfinement transition that resembles some aspects of the counterpart at strong coupling such as the $N$-dependence of entropy~\cite{Sundborg:1999ue,Aharony:2003sx}. Most crucially, deconfinement can be understood as the condensation of long strings both at weak coupling and at strong coupling. However, it was not clear how linear confinement potential can follow from this picture. In the meantime, a series of work on partial deconfinement~\cite{Hanada:2016pwv,Berenstein:2018lrm,Hanada:2018zxn,Hanada:2019czd} led to the discovery of a microscopic mechanism of confinement in the large-$N$ gauge theories~\cite{Hanada:2023rlk,Hanada:2023krw,Hanada:2020uvt} which is essentially the same as Bose-Einstein condensation. 
(See Sec.~\ref{sec:Hanada-Watanabe-review}.) This mechanism is based only on symmetry and is applicable to both weak and strong coupling. Let us call it \textit{Confinement $=$ BEC mechanism}.\footnote{You may prefer $\simeq$ or $\sim$ depending on what you mean by BEC, confinement, or $=$.} The slowly varying Haar-randomness is a consequence of the Confinement $=$ BEC mechanism. Therefore, we understood how the Confinement $=$ BEC mechanism, which is based on a very basic property of gauge theory, leads to linear confinement potential with Casimir scaling.

Confinement $=$ BEC mechanism leads to various phenomena including partial deconfinement and intriguing pattern of global symmetry breaking~\cite{Hanada:2019kue,Hanada:2021ksu}, coexistence of parton-like and string-like degrees of freedom~\cite{Gautam:2022exf},\footnote{
In Ref.~\cite{Gautam:2022exf}, the partially-deconfined phase of the large-$N$ limit of strongly-coupled lattice gauge theory was studied, and linear confinement potential in the confined sector was observed. We expect the same pattern in the continuum theory, by applying the same reasoning used in this paper to the confined sector. Specifically, the Polyakov line restricted to the confined sector in the partially-deconfined phase should be slowly varying Haar random. 
} instanton condensation~\cite{Hanada:2023rlk,Hanada:2023krw}, and so on. Hopefully, this mechanism leads us to a unified picture of an even richer list of nontrivial phenomena. 
\begin{center}
\Large{\textbf{Acknowledgement}}
\end{center}
The authors would like to thank Pavel Buividovich, Nadav Drukker, Jeff Greensite, Jack Holden, Enrico Rinaldi, Andreas Sch\"{a}fer, Hidehiko Shimada, Yuya Tanizaki, Masaki Tezuka, and Hiromasa Watanabe.
V.~G. thanks STFC for the Doctoral Training Programme funding (ST/W507854-2021 Maths DTP).
M.~H. thanks his STFC consolidated grant ST/X000656/1. 
G.~B.\ is funded by the Deutsche Forschungsgemeinschaft (DFG) under Grant No.~432299911 and 431842497.
\appendix
\section{Path integral for Two-dimensional pure Yang-Mills}\label{appendix:2dYM}
\hspace{0.51cm}
In this appendix, we explain how the mechanism we discussed in the main text works for the simple yet illuminating case of pure Yang-Mills theory in two dimensions. Unlike the main text, we consider the spacetime lattice and Euclidean path integral~\cite{Gross:1980he,Wadia:2012fr}. Note that we are merely rewriting known facts by using slightly different language so that the connection to the discussion in the main text becomes manifest. 

Wilson's plaquette action is given by 
\begin{align}
    S_{\rm Wilson} = -\sum_{\vec{n}} \frac{1}{2g^2a^2} \Tr{U_{\Box,\vec{n}} + U_{\Box,\vec{n}}^\dagger}\, , 
\end{align}
where the plaquette term $U_{\Box,\vec{n}}$ is 
\begin{align}
    U_{\Box,\vec{n}} = U_{\vec{n},t} U_{\vec{n}+\hat{t},x} U_{\vec{n}+\hat{x},t}^\dagger U_{\vec{n},x}^\dagger\, . 
\end{align}
We take the spatial direction to be noncompact. (We allow the temporal direction to be compact because we want to consider finite temperature.) Then, we can choose the gauge such that $U_{\vec{n},x }=\textbf{1}$ for all $\vec{n}$, or in other words $A_x = 0$. In this gauge, the plaquette term reduces to $U_{\Box,\vec{n}} = U_{\vec{n},t} U^\dagger_{\vec{n}+\hat{x},t}$. 
The partition function is 
\begin{align}
    Z 
    = \int \left(\prod_{\vec{n}}dU_{\vec{n},t}dU_{\vec{n},x}\right)\ e^{-S_{\rm Wilson}[U_t,U_x]}
    = \int \left(\prod_{\vec{n}}dU_{\vec{n},t}\right)\ e^{-S_{\rm Wilson}[U_t,U_x=\textbf{1}]}\, . 
    \label{eq:partition_function_2d_v1}
\end{align}
The integral is taken over the Haar measure. 
To see that no extra factor appears, we can write the spatial link variables as $U_{\vec{n},x}=\Omega_{\vec{n}}^\dagger\Omega_{\vec{n}+\hat{x}}$ (it is possible because the spatial dimension is noncompact), and use the gauge invariance of the action and the property of Haar measure as follows. 
First, because of the gauge invariance of the action, 
\begin{align}
S_{\rm Wilson}[U_t,U_x]
=
S_{\rm Wilson}[U'_t,U'_x=\textbf{1}]\, ,
\end{align}
where
\begin{align}
U'_{\vec{n},t}
\equiv
\Omega_{\vec{n}}
U_{\vec{n},t}
\Omega_{\vec{n}+\hat{t}}^\dagger\, , 
\quad
U'_{\vec{n},x}
\equiv
\Omega_{\vec{n}}
U_{\vec{n},x}
\Omega_{\vec{n}+\hat{x}}^\dagger\, . 
\end{align}
The Haar measure is also gauge invariant, and hence, 
\begin{align}
    Z 
    = \int \left(\prod_{\vec{n}}dU'_{\vec{n},t}dU'_{\vec{n},x}\right)\ e^{-S_{\rm Wilson}[U'_t,U'_x=\textbf{1}]}\, . 
\end{align}
The integral over $U'_{\vec{n},x}$ gives a trivial volume factor. This is the case not just for partition function but also for evaluation of gauge-invariant observables. Therefore, we drop $dU'_{\vec{n},x}$, and removing the prime, we arrive at \eqref{eq:partition_function_2d_v1}.

Redefining the variables such that $W_{\vec{n}} \equiv U_{\vec{n},t} U^\dag_{\vec{n}+\hat{x},t} = U_{\Box,\vec{n}}$, the action reduces simply to
\begin{align}
    S_{\rm Wilson} = -\sum_{\vec{n}} \frac{1}{2g^2a^2} \Tr{W_{\Vec{n}} + W^\dag_{\Vec{n}}}\, .  
\end{align}
The partition function is 
\begin{align}
    Z 
    &= \int\left(\prod_{\vec{n}}dW_{\Vec{n}}\right)\exp{\sum_{\vec{n}} \frac{1}{2g^2a^2} \Tr{W_{\Vec{n}} + W^\dag_{\Vec{n}}}}
    \nonumber\\
    &= \prod_{\vec{n}}\left(\int dW_{\Vec{n}} \; \exp{\frac{1}{2g^2a^2} \Tr{W_{\Vec{n}} + W^\dag_{\Vec{n}}}}\right)\, . 
\end{align}
Again, there is no extra factor due to the property of the Haar measure. From this expression, we see that plaquettes $W_{\vec{n}}$ behave as independent variables.

\begin{figure}[htbp]
\begin{center}
\scalebox{0.25}{
\includegraphics{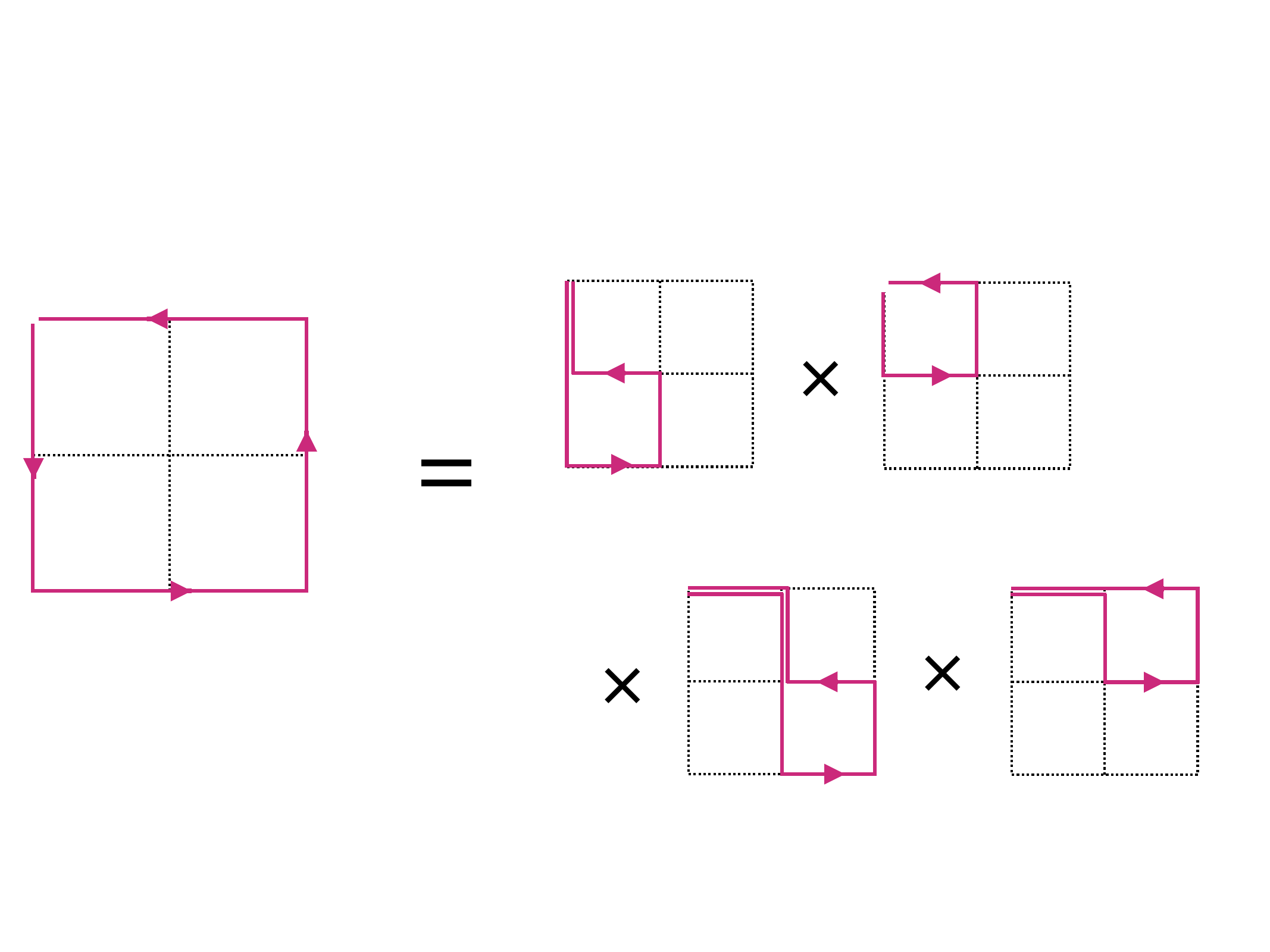}}
\end{center}
\caption{
Wilson loop expressed in terms of a product of plaquettes with `tails'.
}\label{fig:Wilson_prod}
\end{figure}

The expectation value of a Wilson loop $\mathrm{Tr}\mathcal{W}_C$ along a contour $C$ can then be calculated as 
\begin{align}
    \langle \mathrm{Tr}\mathcal{W}_C \rangle = \frac{1}{Z} \int dW \mathcal{W}_C e^{-S_{\text{Wilson}}}
\end{align}
Here, we use $\mathcal{W}_C$ to denote the loop before taking a trace, which is an $N\times N$ matrix for SU($N$) theory. 
Any Wilson loop $\mathcal{W}_C$ can be broken down in terms of the plaquettes contained in it, as shown in Fig.~\ref{fig:Wilson_prod}. More precisely, Wilson loops can be expressed as a product of plaquettes with `tails'. 
Specifically in two dimensions, each plaquette $W_{\Vec{n}}$ is independent of the others, and hence, the plaquette with `tail' is also independent.
Therefore, the Wilson loop in two dimensions factorizes to the product of plaquettes contained in it. Hence, for a Wilson loop consisting of $n_{\rm plaq.}$ plaquettes, we obtain
\begin{align}
    \langle \mathcal{W}_C\rangle 
    = 
    \left(w\cdot\textbf{1}\right)^{n_{\rm plaq.}}
    = 
    w^{n_{\rm plaq.}}\cdot\textbf{1}\, , 
\end{align}
where $w$ is the expectation value of single plaquette, i.e. $\langle W_{\Vec{n}} \rangle=w\cdot\textbf{1}$. 

When the lattice spacing $a$ is small, $W_{\vec{n}}$ localizes around $\textbf{1}$. Let us introduce a traceless Hermitian matrix $X_{\vec{n}}$ by 
\begin{align}
W_{\vec{n}}=e^{iaX_{\vec{n}}}\, . 
\end{align}
By construction, this $X$ is the same as field strength.
Then, 
\begin{align}
    S_{\rm Wilson} 
    =
    \frac{1}{2g^2}
    \sum_{\vec{n}}\mathrm{Tr}X^2_{\vec{n}}
    +
    O(a^2)\, .  
\end{align}
The Haar measure $dW$ is equivalent to the flat measure $dX$. Therefore, $X_{\vec{n}}$ is Gaussian random and the variance is $g^2$. 
The thin loop of the form shown in Fig.~\ref{fig:Figure_W}, with one-lattice-unit extension along the spatial direction, is obtained by taking a product of $n_t$ plaquettes with tails, where $n_t$ is the number of lattice sites along the temporal direction which is related to temperature by $\beta=T^{-1}=an_t$. The sum of $n_t$ independent Gaussian random numbers with variance $g^2$ is a Gaussian random number with variance $g^2n_t$. Therefore, the thin loop takes the form of $e^{iga\sqrt{n_t}\tilde{X}}=e^{ig\sqrt{\beta a}\tilde{X}}$, where $\tilde{X}$ is Gaussian random with variance 1. This corresponds to $V_n^{-1}V_{n+1}$ in Sec.~\ref{sec:epsilon-estimate}. Because all plaquettes are independent, $V_n^{-1}V_{n+1}$'s are also independent.

\bibliographystyle{utphys}
\bibliography{RMP-Confinement}

\providecommand{\href}[2]{#2}\begingroup\raggedright\begin{thebibliography}{10}

\bibitem{Hanada:2023rlk}
M.~Hanada and H.~Watanabe, ``{On thermal transition in QCD},''
  \href{http://arxiv.org/abs/2310.07533}{{\ttfamily arXiv:2310.07533
  [hep-th]}}.

\bibitem{Hanada:2023krw}
M.~Hanada, H.~Ohata, H.~Shimada, and H.~Watanabe, ``{A new perspective on
  thermal transition in QCD},''
  \href{http://arxiv.org/abs/2310.01940}{{\ttfamily arXiv:2310.01940
  [hep-th]}}.

\bibitem{Hanada:2020uvt}
M.~Hanada, H.~Shimada, and N.~Wintergerst, ``{Color confinement and
  Bose-Einstein condensation},''
  \href{http://dx.doi.org/10.1007/JHEP08(2021)039}{{\em JHEP} {\bfseries 08}
  (2021) 039}, \href{http://arxiv.org/abs/2001.10459}{{\ttfamily
  arXiv:2001.10459 [hep-th]}}.

\bibitem{Ambjorn:1984mb}
J.~Ambjorn, P.~Olesen, and C.~Peterson, ``{Stochastic Confinement and
  Dimensional Reduction. 1. Four-Dimensional SU(2) Lattice Gauge Theory},''
  \href{http://dx.doi.org/10.1016/0550-3213(84)90475-9}{{\em Nucl. Phys. B}
  {\bfseries 240} (1984) 189--212}.

\bibitem{Ambjorn:1984dp}
J.~Ambjorn, P.~Olesen, and C.~Peterson, ``{Stochastic Confinement and
  Dimensional Reduction. 2. Three-dimensional SU(2) Lattice Gauge Theory},''
  \href{http://dx.doi.org/10.1016/0550-3213(84)90242-6}{{\em Nucl. Phys. B}
  {\bfseries 240} (1984) 533--542}.

\bibitem{DelDebbio:1995gc}
L.~Del~Debbio, M.~Faber, J.~Greensite, and S.~Olejnik, ``{Casimir scaling
  versus Abelian dominance in QCD string formation},''
  \href{http://dx.doi.org/10.1103/PhysRevD.53.5891}{{\em Phys. Rev. D}
  {\bfseries 53} (1996) 5891--5897},
  \href{http://arxiv.org/abs/hep-lat/9510028}{{\ttfamily
  arXiv:hep-lat/9510028}}.

\bibitem{Bali:2000un}
G.~S. Bali, ``{Casimir scaling of SU(3) static potentials},''
  \href{http://dx.doi.org/10.1103/PhysRevD.62.114503}{{\em Phys. Rev. D}
  {\bfseries 62} (2000) 114503},
  \href{http://arxiv.org/abs/hep-lat/0006022}{{\ttfamily
  arXiv:hep-lat/0006022}}.

\bibitem{Deldar:1999vi}
S.~Deldar, ``{Static SU(3) potentials for sources in various
  representations},'' \href{http://dx.doi.org/10.1103/PhysRevD.62.034509}{{\em
  Phys. Rev. D} {\bfseries 62} (2000) 034509},
  \href{http://arxiv.org/abs/hep-lat/9911008}{{\ttfamily
  arXiv:hep-lat/9911008}}.

\bibitem{Gupta:2007ax}
S.~Gupta, K.~Huebner, and O.~Kaczmarek, ``{Renormalized Polyakov loops in many
  representations},'' \href{http://dx.doi.org/10.1103/PhysRevD.77.034503}{{\em
  Phys. Rev. D} {\bfseries 77} (2008) 034503},
  \href{http://arxiv.org/abs/0711.2251}{{\ttfamily arXiv:0711.2251 [hep-lat]}}.

\bibitem{Gupta:2006qm}
S.~Gupta, K.~Huebner, and O.~Kaczmarek, ``{Polyakov loop in different
  representations of SU(3) at finite temperature},''
  \href{http://dx.doi.org/10.1016/j.nuclphysa.2006.11.160}{{\em Nucl. Phys. A}
  {\bfseries 785} (2007) 278--281},
  \href{http://arxiv.org/abs/hep-lat/0608014}{{\ttfamily
  arXiv:hep-lat/0608014}}.

\bibitem{Mykkanen:2012ri}
A.~Mykkanen, M.~Panero, and K.~Rummukainen, ``{Casimir scaling and
  renormalization of Polyakov loops in large-N gauge theories},''
  \href{http://dx.doi.org/10.1007/JHEP05(2012)069}{{\em JHEP} {\bfseries 05}
  (2012) 069}, \href{http://arxiv.org/abs/1202.2762}{{\ttfamily arXiv:1202.2762
  [hep-lat]}}.

\bibitem{Brzoska:2004pi}
A.~M. Brzoska, F.~Lenz, J.~W. Negele, and M.~Thies, ``{Diffusion of Wilson
  loops},'' \href{http://dx.doi.org/10.1103/PhysRevD.71.034008}{{\em Phys. Rev.
  D} {\bfseries 71} (2005) 034008},
  \href{http://arxiv.org/abs/hep-th/0412003}{{\ttfamily arXiv:hep-th/0412003}}.

\bibitem{Arcioni:2005iq}
G.~Arcioni, S.~de~Haro, and P.~Gao, ``{A Diffusion model for SU(N) QCD
  screening},'' \href{http://dx.doi.org/10.1103/PhysRevD.73.074508}{{\em Phys.
  Rev. D} {\bfseries 73} (2006) 074508},
  \href{http://arxiv.org/abs/hep-th/0511213}{{\ttfamily arXiv:hep-th/0511213}}.

\bibitem{Buividovich:2006yj}
P.~V. Buividovich and V.~I. Kuvshinov, ``{Kramers-Moyall cumulant expansion for
  the probability distribution of parallel transporters in quantum gauge
  fields},'' \href{http://dx.doi.org/10.1103/PhysRevD.73.094015}{{\em Phys.
  Rev. D} {\bfseries 73} (2006) 094015},
  \href{http://arxiv.org/abs/hep-th/0605207}{{\ttfamily arXiv:hep-th/0605207}}.

\bibitem{Buividovich:2007xh}
P.~V. Buividovich and M.~I. Polikarpov, ``{Random walks of Wilson loops in the
  screening regime},''
  \href{http://dx.doi.org/10.1016/j.nuclphysb.2007.08.013}{{\em Nucl. Phys. B}
  {\bfseries 790} (2008) 28--41},
  \href{http://arxiv.org/abs/0704.3367}{{\ttfamily arXiv:0704.3367 [hep-ph]}}.

\bibitem{Poppitz:2017ivi}
E.~Poppitz and M.~E. Shalchian~T., ``{String tensions in deformed Yang-Mills
  theory},'' \href{http://dx.doi.org/10.1007/JHEP01(2018)029}{{\em JHEP}
  {\bfseries 01} (2018) 029}, \href{http://arxiv.org/abs/1708.08821}{{\ttfamily
  arXiv:1708.08821 [hep-th]}}.

\bibitem{Hanada:2016pwv}
M.~Hanada and J.~Maltz, ``{A proposal of the gauge theory description of the
  small Schwarzschild black hole in AdS$_5\times$S$^5$},''
  \href{http://dx.doi.org/10.1007/JHEP02(2017)012}{{\em JHEP} {\bfseries 02}
  (2017) 012}, \href{http://arxiv.org/abs/1608.03276}{{\ttfamily
  arXiv:1608.03276 [hep-th]}}.

\bibitem{Berenstein:2018lrm}
D.~Berenstein, ``{Submatrix deconfinement and small black holes in AdS},''
  \href{http://dx.doi.org/10.1007/JHEP09(2018)054}{{\em JHEP} {\bfseries 09}
  (2018) 054}, \href{http://arxiv.org/abs/1806.05729}{{\ttfamily
  arXiv:1806.05729 [hep-th]}}.

\bibitem{Hanada:2018zxn}
M.~Hanada, G.~Ishiki, and H.~Watanabe, ``{Partial Deconfinement},''
  \href{http://dx.doi.org/10.1007/JHEP03(2019)145}{{\em JHEP} {\bfseries 03}
  (2019) 145}, \href{http://arxiv.org/abs/1812.05494}{{\ttfamily
  arXiv:1812.05494 [hep-th]}}. [Erratum: JHEP 10, 029 (2019)].

\bibitem{Hanada:2019czd}
M.~Hanada, A.~Jevicki, C.~Peng, and N.~Wintergerst, ``{Anatomy of
  Deconfinement},'' \href{http://dx.doi.org/10.1007/JHEP12(2019)167}{{\em JHEP}
  {\bfseries 12} (2019) 167}, \href{http://arxiv.org/abs/1909.09118}{{\ttfamily
  arXiv:1909.09118 [hep-th]}}.

\bibitem{Luscher:2010iy}
M.~L\"uscher, ``{Properties and uses of the Wilson flow in lattice QCD},''
  \href{http://dx.doi.org/10.1007/JHEP08(2010)071}{{\em JHEP} {\bfseries 08}
  (2010) 071}, \href{http://arxiv.org/abs/1006.4518}{{\ttfamily arXiv:1006.4518
  [hep-lat]}}. [Erratum: JHEP 03, 092 (2014)].

\bibitem{Datta:2015bzm}
S.~Datta, S.~Gupta, and A.~Lytle, ``{Using Wilson flow to study the SU(3)
  deconfinement transition},''
  \href{http://dx.doi.org/10.1103/PhysRevD.94.094502}{{\em Phys. Rev. D}
  {\bfseries 94} no.~9, (2016) 094502},
  \href{http://arxiv.org/abs/1512.04892}{{\ttfamily arXiv:1512.04892
  [hep-lat]}}.

\bibitem{Petreczky:2015yta}
P.~Petreczky and H.~P. Schadler, ``{Renormalization of the Polyakov loop with
  gradient flow},'' \href{http://dx.doi.org/10.1103/PhysRevD.92.094517}{{\em
  Phys. Rev. D} {\bfseries 92} no.~9, (2015) 094517},
  \href{http://arxiv.org/abs/1509.07874}{{\ttfamily arXiv:1509.07874
  [hep-lat]}}.

\bibitem{DiGiacomo:2000irz}
A.~Di~Giacomo, H.~G. Dosch, V.~I. Shevchenko, and Y.~A. Simonov, ``{Field
  correlators in QCD: Theory and applications},''
  \href{http://dx.doi.org/10.1016/S0370-1573(02)00140-0}{{\em Phys. Rept.}
  {\bfseries 372} (2002) 319--368},
  \href{http://arxiv.org/abs/hep-ph/0007223}{{\ttfamily arXiv:hep-ph/0007223}}.

\bibitem{Greensite:2006sm}
J.~Greensite, K.~Langfeld, S.~Olejnik, H.~Reinhardt, and T.~Tok, ``{Color
  Screening, Casimir Scaling, and Domain Structure in G(2) and SU(N) Gauge
  Theories},'' \href{http://dx.doi.org/10.1103/PhysRevD.75.034501}{{\em Phys.
  Rev. D} {\bfseries 75} (2007) 034501},
  \href{http://arxiv.org/abs/hep-lat/0609050}{{\ttfamily
  arXiv:hep-lat/0609050}}.

\bibitem{Migdal:1975zg}
A.~A. Migdal, ``{Recursion Equations in Gauge Theories},'' {\em Sov. Phys.
  JETP} {\bfseries 42} (1975) 413.

\bibitem{Witten:1991we}
E.~Witten, ``{On quantum gauge theories in two-dimensions},''
  \href{http://dx.doi.org/10.1007/BF02100009}{{\em Commun. Math. Phys.}
  {\bfseries 141} (1991) 153--209}.

\bibitem{Nguyen:2021naa}
M.~Nguyen, Y.~Tanizaki, and M.~\"Unsal, ``{Noninvertible 1-form symmetry and
  Casimir scaling in 2D Yang-Mills theory},''
  \href{http://dx.doi.org/10.1103/PhysRevD.104.065003}{{\em Phys. Rev. D}
  {\bfseries 104} no.~6, (2021) 065003},
  \href{http://arxiv.org/abs/2104.01824}{{\ttfamily arXiv:2104.01824
  [hep-th]}}.

\bibitem{Sundborg:1999ue}
B.~Sundborg, ``{The Hagedorn transition, deconfinement and N=4 SYM theory},''
  \href{http://dx.doi.org/10.1016/S0550-3213(00)00044-4}{{\em Nucl. Phys. B}
  {\bfseries 573} (2000) 349--363},
  \href{http://arxiv.org/abs/hep-th/9908001}{{\ttfamily arXiv:hep-th/9908001}}.

\bibitem{Aharony:2003sx}
O.~Aharony, J.~Marsano, S.~Minwalla, K.~Papadodimas, and M.~Van~Raamsdonk,
  ``{The Hagedorn - deconfinement phase transition in weakly coupled large N
  gauge theories},'' \href{http://dx.doi.org/10.4310/ATMP.2004.v8.n4.a1}{{\em
  Adv. Theor. Math. Phys.} {\bfseries 8} (2004) 603--696},
  \href{http://arxiv.org/abs/hep-th/0310285}{{\ttfamily arXiv:hep-th/0310285}}.

\bibitem{Hanada:2019kue}
M.~Hanada and B.~Robinson, ``{Partial-Symmetry-Breaking Phase Transitions},''
  \href{http://dx.doi.org/10.1103/PhysRevD.102.096013}{{\em Phys. Rev. D}
  {\bfseries 102} no.~9, (2020) 096013},
  \href{http://arxiv.org/abs/1911.06223}{{\ttfamily arXiv:1911.06223
  [hep-th]}}.

\bibitem{Hanada:2021ksu}
M.~Hanada, J.~Holden, M.~Knaggs, and A.~O'Bannon, ``{Global symmetries and
  partial confinement},'' \href{http://dx.doi.org/10.1007/JHEP03(2022)118}{{\em
  JHEP} {\bfseries 03} (2022) 118},
  \href{http://arxiv.org/abs/2112.11398}{{\ttfamily arXiv:2112.11398
  [hep-th]}}.

\bibitem{Gautam:2022exf}
V.~Gautam, M.~Hanada, J.~Holden, and E.~Rinaldi, ``{Linear confinement in the
  partially-deconfined phase},''
  \href{http://dx.doi.org/10.1007/JHEP03(2023)195}{{\em JHEP} {\bfseries 03}
  (2023) 195}, \href{http://arxiv.org/abs/2208.14402}{{\ttfamily
  arXiv:2208.14402 [hep-th]}}.

\bibitem{Gross:1980he}
D.~J. Gross and E.~Witten, ``{Possible Third Order Phase Transition in the
  Large N Lattice Gauge Theory},''
  \href{http://dx.doi.org/10.1103/PhysRevD.21.446}{{\em Phys. Rev. D}
  {\bfseries 21} (1980) 446--453}.

\bibitem{Wadia:2012fr}
S.~R. Wadia, ``{A Study of U(N) Lattice Gauge Theory in 2-dimensions},''
  \href{http://arxiv.org/abs/1212.2906}{{\ttfamily arXiv:1212.2906 [hep-th]}}.

\end{thebibliography}\endgroup

\end{document}